\documentclass{statement}

\usepackage{palatino}  
\usepackage[T1]{fontenc}
\definecolor{linkcol}{rgb}{0.278,0.541,0.459}%

\definecolor{sectcol}{rgb}{0.63,0.16,0.16}
\definecolor{palevioletred4}{rgb}{0.545,0.278,0.365}
\definecolor{gray40}{rgb}{0.40,0.40,0.40}
\definecolor{gray26}{rgb}{0.26,0.26,0.26}
\definecolor{olive}{rgb}{0.5,0.5,0.0}
\definecolor{gray78}{cmyk}{0,0,0,0.22}

\usepackage[
    pdftitle={Terminologies for Reproducible Research},
    pdfauthor={Lorena Barba},
    pdfpagemode={UseOutlines},
    pdfpagelayout={TwoColumnRight},
    bookmarks, bookmarksopen,bookmarksnumbered={True},
    pdfstartview={FitH},
    breaklinks=true,
    colorlinks, linkcolor={linkcol},citecolor={linkcol},urlcolor={linkcol}
    ]{hyperref}

\makeatletter
\def\url@leostyle{%
  \@ifundefined{selectfont}{\def\UrlFont{\sf}}{\def\UrlFont{\footnotesize\ttfamily}}}
\makeatother
\def\rrr#1\\{\par
\medskip\hbox{\vbox{\parindent=2em\hsize=6.12in
\hangindent=4em\hangafter=1#1}}}

\newlength{\up}
\newlength{\hup}
\begin{document}

\pagenumbering{arabic}
\renewcommand{\thepage} {\arabic{page}}

\thispagestyle{empty}

{ \Huge Terminologies for Reproducible Research} 
\medskip

Lorena A. Barba, the George Washington University, Washington D.C. 

January 2018

\vspace{1cm}

\section*{Introduction}
\vspace{\up}

Reproducible research---by its many names---has come to be regarded as a key concern across disciplines and stakeholder groups. Funding agencies and journals, professional societies and even mass media are paying attention, often focusing on the so-called ``crisis'' of reproducibility. One big problem keeps coming up among those seeking to tackle the issue: different groups are using terminologies in utter contradiction with each other. 
In July 2017, over a dozen participants joined the \href{https://collegeville.github.io/repeto/ReproducibilityWorkshop2017.html}{Workshop on Reproducibility Taxonomies for Computing and Computational Science} at the National Science Foundation, to appraise the variety of terminologies.
My presentation at that event \cite[]{barba2017-repeto} condensed a catalog of uses of the recurrent terms \emph{reproduce} and \emph{replicate}, often meaning different things but sometimes interchangeable.

Looking at a broad sample of publications in different fields, we can classify their terminology via decision tree: they either, \emph{A}---make no distinction between the words \emph{reproduce} and \emph{replicate}, or \emph{B}---use them distinctly.  
If \emph{B}, then they are commonly divided in two camps. 
In a spectrum of concerns that starts at a minimum standard of ``same data$+$same methods$=$same results,'' to ``new data and/or new methods in an independent study$=$same findings,'' group $1$ calls the minimum standard \emph{reproduce}, while group $2$ calls it \emph{replicate}.
This direct swap of the two terms aggravates an already weighty issue.
By attempting to inventory the terminologies across disciplines, I hope that some patterns will emerge to help us resolve the contradictions.

\section*{Pioneers of reproducible research}
\vspace{\up}

The first appearance of the phrase ``reproducible research'' in a scholarly publication appears to be an invited paper presented at the 1992 meeting of the Society of Exploration Geophysics (SEG), from the group of Jon Claerbout at Stanford \cite[]{claerbout1992}.
Claerbout pioneered the use of computers in processing and filtering seismic exploration data \cite[]{wiki:claerbout}.
From at least 1990, he required his students' PhD theses to conform to a standard of reproducibility.
His idea of reproducible research was to leave finished work (an article or a thesis) in a state that allowed colleagues to reproduce the calculation, analysis and final figures by executing a single command.
The goal was to merge a publication with its underlying computational analysis.
They used an automation tool called \texttt{make}, which builds software from source code by reading through a \texttt{makefile}: a list of commands to be executed in sequence.
Their workflow combined a set of standardized commands (burn, build, view, clean), and a filing system for the research compendium associated with the paper (data sets, programs, scripts, parameter files, makefiles).
Explicitly, \cite{claerbout1992} provide a ``definition of reproducibility in computationally oriented research.''
The original SEG paper is somewhat dated, but the group presented an updated overview in \cite{schwabETal2000}.
Here, the authors emphasize the limitations of the traditional methods of scientific publication, especially for computational research. 
In their vision of reproducible research, readers should be able to rebuild published results ``using the author's underlying programs and raw data.''
Implicitly, they are advocating for open code and data.

At Stanford, statistics professor David Donoho learned of Claerbout's methods in the early 1990s, and began adopting (and later promoting) them. 
In \cite{buckheit_donoho1995}, the authors state that ``reproducibility of experiments in seismic exploration requires having the complete software environment available in other laboratories and the full source code available for inspection, modification, and application under varied parameter settings.'' 
This is because the goal in this field is to produce an image of the earth's sub-surface, from the datasets of seismic exploration signals. The software generating the image is key to the final results. 
And here is where the authors write the often-quoted ``slogan'' of reproducible research (paraphrasing Claerbout): 
\begin{quote}
\emph{``An article about computational science in a scientific publication is not the scholarship itself, it is merely advertising of the scholarship. The actual scholarship is the complete software development environment and the complete set of instructions which generated the figures.''}
\end{quote}

\cite{buckheit_donoho1995} include many technical details that seem dated now, so perhaps a better place to read about this group's thinking and practice is \cite{donohoETal2009}. 
This appears to be the first article to publicly state that reproducibility depends on open code and data. The authors define \emph{reproducible computational research} as that ``in which all details of computations---code and data---are made conveniently available to others.'' 
They used the now-recurrent charged language about a ``credibility crisis'' in computational science, and worry that computation cannot claim to be the ``third branch'' of science because most computational results cannot be verified. 
In the two traditional branches, standards of practice already exist for managing the ubiquity of error: deductive science uses formal logic and the mathematical proof, while empirical science uses statistical hypothesis testing and detailed methods reporting. 
The authors also counter the view that reproducibility means re-implementing the research software from the ground up. 
Because, in that case, if the new results did not match, ``the only way we'd ever get to the bottom of such a discrepancy is if we both worked reproducibly and studied detailed differences between code and data.''

The pioneering efforts of Claerbout and Donoho influenced many others. But their concerns centered on research involving computational analysis of unique data (recordings of seismic waves). 
They did not deal with the situation where another researcher might collect \emph{new} data, re-doing a study or an experiment following an original design, then analyze this data to compare the final findings with the original work. 
Situations like this arise in many empirical fields, where scientific findings must be confirmed by independent studies. 
\cite{pengETal2006} distinguish the term \emph{replication} for this scenario. They say:
``Scientific evidence is strengthened when important findings are \emph{replicated} by multiple independent investigators using independent data, analytical methods, laboratories, and instruments.'' 
But epidemiologic studies are expensive, time-consuming, and often impossible to replicate (e.g., when they rely on opportunistic data, say, from an infectious outbreak). 
``An attainable minimum standard is \emph{reproducibility}, which calls for data sets and software to be made available for verifying published findings and conducting alternative analyses,'' they say.

In the social sciences, Harvard professor Gary King, one of the most cited political scientists of this generation, also pioneered the reproducibility concerns. 
In \cite{king1995}, he writes: ``The replication standard holds that sufficient information exists with which to understand, evaluate, and build upon a prior work if a third party could replicate the results without any additional information from the author.'' He uses the term \emph{replication} throughout, not making a distinction like we have in the Claerbout/Donoho/Peng writings on reproducible research. 
In fact, his paper uses ``replication'' 96 times, ``replicate'' 21 times, and ``reproduce'' (as verb) only twice. E.g., ``good science requires that we be able to reproduce existing numerical results\ldots''
A more recent work co-authored by him \cite[]{lazerETal2014}, still does not use the words ``reproduce/reproducibility/reproducible,'' and only uses ``replicate.'' 
This appears to be the common usage in his field, as a Google Scholar search with ``replication political science`` gives 339,000 results, and with ``reproducible political science'' gives just 59,700 results (checked 01/14/2017).

\section*{Conflicting terminologies}
\vspace{\up}

As mentioned above, Claerbout and Donoho's pioneering work influenced many others, who adopt their definition of reproducible research. 
For example, \cite{gentlemanETal2007}: ``By reproducible research, we mean research papers with accompanying software tools that allow the reader to directly reproduce the results and employ the computational methods that are presented in the research paper.''
\cite{vandewalleETal2009}: ``A research work is called reproducible if all information relevant to the work, including, but not limited to, text, data and code, is made available, such that an independent researcher can reproduce the results.''
\cite{leveque2009}: ``The idea of `reproducible research' in scientific computing is to archive and make publicly available all the codes used to create a paper's figures or tables, preferably in such a manner that readers can download the codes and run them to reproduce the results.''
The distinction with the idea of replication, introduced by Peng, also appears in other works. 
The \emph{Annals of Internal Medicine} issued a statement saying: ``Independent replication by independent scientists in independent settings provides the best assurance that a scientific finding is valid; however, the resources and time required for high-quality clinical studies makes literal replication of published studies a slow corrective to any errors in the original publication. However, scientists and journal editors can promote `reproducible research,' [which ensures] that independent scientists can reproduce published results by using the same procedures and data as the original investigators'' \cite[]{laineETal2007}. 
In an editorial, the \emph{Int. Journal of Forecasting} announces that it will publish \emph{replication studies}: works attempting to independently verify research findings ``under the same or very similar conditions'' \cite[]{hyndman2010}. 
\emph{Science} published a special issue on ``Data Replication \& Reproducibility'' in 2011, to which the introduction reads: ``Replication---the confirmation of results and conclusions from one study obtained independently in another---is considered the scientific gold standard'' \cite[]{jasnyETal2011}.
Peng's own (widely cited) article in this special issue introduces the idea of a \emph{reproducibility spectrum}, in which reproducible research is a ``minimum standard for judging scientific claims when full independent replication of a study is not possible.'' \cite[]{peng2011}

On the basis of the multiple references cited above, here are concise definitions that convey the meanings of the Claerbout/Donoho/Peng convention:

\begin{description}
\item[Reproducible research:] Authors provide all the necessary data and the computer codes to run the analysis again, re-creating the results.
\item[Replication:] A study that arrives at the same scientific findings as another study, collecting new data (possibly with different methods) and completing new analyses.
\end{description}

Unfortunately, conflicting terminologies started to spread in recent years. 
It appears that the first work to directly swap the usage of \emph{reproducible} and \emph{replicable} is \cite{drummond2009}. 
This is a workshop contribution to a conference in the field of machine learning (meaning, it received light or no peer review). 
The author seems to arbitrarily swap existing terminology; he says: ``I have used the term replicability for what others have called reproducibility in our literature.''
He provides no more justification than ``I think it reasonable.''
Mark Liberman, Distinguished Professor of Linguistics at the University of Pennsylvania, analyzed the usage of terms in the literature, and referring to Drummond's paper he concluded: ``Since the technical term `reproducible research' has been in use since 1990, and the technical distinction between reproducible and replicable at least since 2006, we should reject [the] attempt to re-coin technical terms reproducible and replicable in senses that assign the terms to concepts nearly opposite to those used in the definitions by Claerbout, Peng and others.'' \cite[]{liberman2015}

Alas, before the scrutiny of the linguistics professor, others picked up on the swapped terminology. 
\cite{casadevall_fang2010}, citing Drummond, adopt it and discuss various scenarios in the fields of microbiology and immunology. 
Other examples include \cite{davison2012}, \cite{loscalzo2012}, \cite{crookETal2013}, \cite{cooper2015}, among many others. 
Drummond's 2009 paper has 134 citations in Google Scholar (checked Jan.\ 15, 2018), although some are for other reasons \cite[]{dewinter_happee2013} or to plainly point out the discrepant terminology \cite[]{boylan2015}.

Some other works adopted the swapped terms without a clear link to \cite{drummond2009}. 
In \cite{levequeETal2012}, referring to topics of discussion at a July 2011 workshop titled ``Reproducible Research: Tools and Strategies for Scientific Computing,'' the authors write: ``As an example of the lack of a common nomenclature, two sequential speakers provided opposite definitions for replicable and reproducible. (We believe the  first refers to the ability to run a code and produce exactly the same results as published, and the second refers to the ability to create a code that independently verifies the published results using the information provided.)'' 
A citation follows to \cite{stodden2011}, where we read: ``Replication, using author-provided code and data, and independent reproduction work hand-in-hand. We can reserve the term `replicability' for the regeneration of published results from author-provided code and data.'' Followed by an immediate citation to \cite{king1995}, maybe this was prompted by the similarity between King's conditions for a ``replication standard'' and the Claerbout/Donoho concept of reproducible research. 
(Although, as mentioned above, King simply uses ``replication'' for everything.) 
Also in \cite{levequeETal2012}, the practice of ``private reproducibility'' is explained as being able to rebuild own past research results from the precise version of the code that was used to create them. 
In this sense, ``reproducibility'' is that minimum standard in the spectrum, in contrast to ``replicability,'' as Drummond would have it. 
The opposing terminologies coexist in this paper.

A more recent use of terminology in contradiction with the Claerbout/Donoho/Peng convention appears in the effort of the Association of Computing Machinery (ACM) on Result and Artifact Review and Badging.\footnote{ Approved June 2016, \url{https://www.acm.org/publications/policies/artifact-review-badging}}  It deploys across ACM publications a badging system for articles complying with various standards of code and data sharing. With that purpose, it defines the following terminology:
\vspace{\up}
\begin{description}
\item[Repeatability ]  (Same team, same experimental setup.)
The measurement can be obtained with stated precision by the same team using the same measurement procedure, the same measuring system, under the same operating conditions, in the same location on multiple trials. For computational experiments, this means that a researcher can reliably repeat her own computation.
\item[Replicability ] (Different team, same experimental setup.)
The measurement can be obtained with stated precision by a different team using the same measurement procedure, the same measuring system, under the same operating conditions, in the same or a different location on multiple trials. For computational experiments, this means that an independent group can obtain the same result using the author?s own artifacts.
\item[Reproducibility] (Different team, different experimental setup.)
The measurement can be obtained with stated precision by a different team, a different measuring system, in a different location on multiple trials. For computational experiments, this means that an independent group can obtain the same result using artifacts which they develop completely independently.
\end{description}

The source cited for the definitions adopted by the ACM is the International Vocabulary of Metrology \cite[]{jcgm2008}, which establishes terminology for physical measurements. 
Relevant passages from this source are (abridged):
\vspace{\up}
\begin{description}
\item[repeatability condition of measurement]\ldots same measurement procedure, same operators, same measuring system, same operating conditions and same location, and replicate measurements on the same or similar objects over a short period of time (p.23);
\item[reproducibility condition of measurement]\ldots set of conditions that includes different locations, operators, measuring systems, and replicate measurements on the same or similar objects (p.24).
\end{description}

The document contains no definition of ``replicability'' but both definitions above use the phrase `replicate measurements.' The scenario for these definitions is a physical quantity being measured, and the precision of that measurement. Repeatability is the precision over successive measurements of the same quantity, with everything kept the same (even the operator), over a short period of time. 
Because measuring tools and procedures have inherent errors, it's relevant to document repeatability in the form of standard deviation or other dispersion characteristics (e.g., graphical).
Reproducibility of measurements involves changing at least one condition, e.g., the instrument, the location, or the operator. 
Because the \emph{same} physical quantity is being measured (``on the same or similar object''), reproducibility of measurement is also expressed by a dispersion statistic, like standard deviation \cite[p.14]{taylorETal1994}. 

Along the way of creating the ACM terminology, based on the metrology document, some leap occurred. Replicability is not defined in \cite{jcgm2008}, and reproducibility could just as well be mapped to the minimum standard of the Peng spectrum, where certainly the operator (researcher) changes, the instrument (computer) does as well, and other conditions may also change, but the object of study remains the same.
Arguably, conditions for physical measurements are a distant analogy for the complex processes of a full scientific workflow. 

\section*{Cataloguing the reproducibility literature}
\vspace{\up}

The conflicting terminologies are at least an annoyance, and at worst an impediment to the progress of science. Yet, no solution is at hand beyond a general good practice of always defining the terms used in any particular writing. 
Here is a decision tree to catalogue the terminologies in the literature: 
authors either, $A$---make no distinction between the words \emph{reproduce} and \emph{replicate}, or $B$---use them distinctly.  
If $B$, then they are commonly divided in two camps. 
In a spectrum of concerns that starts at a minimum standard of ``same data$+$same methods$=$same results,'' to ``new data and/or new methods in an independent study$=$same findings,'' group $1$ calls the minimum standard \emph{reproduce}, while group $2$ calls it \emph{replicate}.
$A$ includes \cite{king1995}.
$B1$ corresponds to the Claerbout/Donoho/Peng convention, while $B2$ agrees with \cite{drummond2009} and the ACM terminology.
Table \ref{repro-table} classifies into these groups all the references cited above, and more.

\bigskip

\begin{table}[h]
\caption{Catalogue of terminologies in the literature, with Google Scholar citations (checked Jan.\ 20, 2018).}
\begin{footnotesize}
\begin{tabular}{c c c}
$A$				&	$B1$				&	$B2$ \\ \hline
\cite{king1995}, 527	& \cite{pengETal2006}, 177	& \cite{drummond2009}, 135 \\
\cite{jcgm2008}, 32	& \cite{gentlemanETal2007}, 216	& \cite{casadevall_fang2010}, 58\\
				& \cite{laineETal2007}, 134	& \cite{stodden2011}, 30\\
\cite{dewaldETal1986}, 506 & \cite{vandewalleETal2009}, 266& \cite{davison2012}, 80 \\
\cite{pesaran2003}, 12 & \cite{leveque2009}, 32	& \cite{loscalzo2012}, 31\\
\cite{mccullough2008}, 93 & \cite{hyndman2010}, 20	& \cite{levequeETal2012}, 74\\
\cite{garijoETal2013}, 52	& \cite{jasnyETal2011}, 180& \cite{crookETal2013}, 16\\
\cite{openscience2012}, 300 & \cite{peng2011}, 552 	& \cite{cooper2015}, 26\\
\cite{openscience2015}, 1573 & 					& \\
\cite{stodden2015}, 19 & \cite{koenkerETal2009}, 58	& \cite{cartwright1991}, 81\\
\cite{duvendackETal2017}, 13 &\cite{delescluseETal2012}, 22 &\cite{pellizzariETal2017}\\
\cite{lejaeghere2016},199	&  \cite{sandveETal2013}, 227	& \cite{faseb2016}\\
\cite{coudert2017}, 3	& \cite{stoddenETal2014}, 119	& \\
				& \cite{topalidou2015}, 14		& \\
				&\cite{iqbalETal2016}, 67		& \\
				& \cite{kafkafiETal2016}, 2		& \\
				&\cite{stevens2017}, 1		& \\
				& \cite{kitzesETal2017}, 10		&\\
				& \cite{benureauETal2017}, 1	& \\
				& \cite{bollenETal2015}, 12		&\\
				& \cite{bromanETal2017}, 4 	&\\
\end{tabular}
\end{footnotesize}
\label{repro-table}
\end{table}%

\subsection*{Added references and commentary}
\vspace{\up}

Table \ref{repro-table} includes additional references to those cited in the narrative above.  Quotes from these and/or commentary to justify their grouping follow below, starting by what seems to be the predominant terminology: that of group $B1$.

\cite{koenkerETal2009} cite the Claerbout `slogan,' from \cite{buckheit_donoho1995}, and list various software tools and techniques to enable reproducible research. 
They then describe two `replication case studies,' i.e., efforts to obtain the same results that appeared in other publications. 
In the conclusion, they say: ``the real challenge of reproducible econometric research lies in restructuring incentives to encourage better archiving and distribution of the gory details of computationally oriented research.''
Despite absence of explicit definitions, the usage agrees with group $B1$. 
Similarly, \cite{delescluseETal2012} cite Claerbout, Donoho, and Peng (among many others), and adopt the $B1$ terminology. 
The authors describe their preferred tools for reproducible data analysis of neurophysiological data, and illustrate the process with a detailed example of implementation.

The work by \cite{sandveETal2013} is also in group $B1$. It says: ``As full replication of studies on independently collected data is often not feasible, [\ldots] reproducible research [is] an attainable minimum standard for assessing the value of scientific claims.''
The authors use the phrase `replication studies' several times,  and refer to `inability to replicate findings' and `replication in studies with different data.'
Thus, although they included no explicit terminology definition, their usage agrees with  Claerbout/Donoho/Peng.

\cite{garijoETal2013} lacks any definitions, and does not seem to make a distinction between \emph{replicate} and \emph{reproduce}. However, the first term is used only 3 times, while terms with the root `reproduc-' appear 161 times. The paper deals with reproducing another group's published work using both artifacts provided by original authors and others.
It's listed in group $A$ as a conservative choice, but it could be argued to belong in $B1$.

The first comprehensive review of the field in book form, edited by \cite{stoddenETal2014}, includes this passage in the preface:
``Replication, the practice of independently implementing scientific experiments to validate specific findings, is the cornerstone of discovering scientific truth. Related to replication is reproducibility, which is the calculation of quantitative scientific results by independent scientists using the original datasets and methods.'' 
This usage agrees with group $B1$, though on inspection of the various chapters, some authors don't make a clear distinction between the terms `reproduce/replicate,` and often use `reproducibility` as an umbrella term. 
An example, from chapter 11 (by the Open Science Collaboration): ``\ldots narrowly, reproducibility is the repetition of a simulation or data analysis of existing data by re-executing a program. More broadly, reproducibility refers to \emph{direct replication}, an attempt to replicate the original observation using the same methods of a previous investigation but collecting new [data].''

\cite{topalidou2015} describe an undertaking to reproduce published results that failed initially when the code provided by the original authors did not compile. 
The effort resulted in a collaboration with the original authors, and a reimplementation of the code from scratch, ``following the principles of reproducible computational science as proposed in \cite{peng2011}.'' 
The language agrees with group $B1$.

In \cite{iqbalETal2016}, we read: ``There are many different proposals on how reproducible research can be guaranteed. These include approaches at reproducible practices, i.e., making other investigators able to repeat the process and calculations; re-analysis (as in the case of randomized trials); and replication by independent investigators, as in genetics, psychology, and cancer biology.''
This usage agrees with group $B1$.

A report of the January 2015 workshop ``Replicability and reproducibility of discoveries in animal phenotyping'' at Tel Aviv University \cite[]{kafkafiETal2016}, which included researchers from many disciplines (genetics, behavior genetics, behavioral neuroscience, ethology, statistics, bioinformatics and database programming), also aligns with the $B1$ terminology. 
It says: ``the scientific community has become increasingly concerned with cases of published `discoveries' that could not be replicated in subsequent studies, and sometimes could not even be reproduced in reanalysis of the original data.'' 
The meeting was hosted by the Replicability Research Group, with funding from the European Research Council. 
On their website,\footnote{\;\url{http://www.replicability.tau.ac.il/index.php/replicability-in-science/replicability-vs-reproducibility.html}} this group notes their terminology usage, citing the policy in the \emph{Biostatistics} journal \cite[]{peng2009}.

\cite{stevens2017} includes explicit definitions: ``Replicability is `re-performing the experiment and collecting new data,' whereas reproducibility is `re-performing the same analysis with the same code using a different analyst.' Therefore, one can replicate a study or an effect (outcome of a study) but reproduce results (data analyses).''
This usage aligns with group $B1$.

The recent book edited by \cite{kitzesETal2017} consists of 31 case studies that present concrete practices for reproducibility in data-intensive research. 
The book also includes synthesis chapters that aim to summarize the lessons learned from the various contributions. 
In the first of these, \cite{marwickETal2017} address definitions, mentioning the confusion that has spread over the recent years, and decidedly adopting the $B1$ terminology. 
They note that this is ``broadly consistent with usage of these terms throughout this book'' (more than 40 authors).

\cite{benureauETal2017} use an example to discuss the necessary characteristics of scientific codes: 
``The code should be executable (re-runnable) and produce the same result more than once (repeatable); it should allow an investigator to reobtain the published results (reproducible) while being easy to use, understand and modify (reusable), and it should act as an available reference for any ambiguity in the algorithmic descriptions of the article (replicable).'' 
Their meanings for `reproducible' and `replicable' are aligned with group $B1$, but they also expand the spectrum of concerns, as it specifically relates to code. 
Using a concrete example lends to clear and illuminating explanations of the potential pitfalls with scientific use of software.

Moving on to group $B2$, with the opposite terminology, we find in \cite{cartwright1991}:
``I propose to distinguish replicability---doing the same experiment again---from reproducibility---doing a new experiment.'' 
This essay by a philosopher of science is a commentary on the work of another, who had proposed that ``replication is the establishment of a new and contested result by agreement over what counts as a correctly performed series of experiments''  \cite[]{collins1991}. 
Both authors are concerned with replication in economics, where usage generally makes no distinction with the term `reproduce,' and most often defaults to the term `replicate'  for everything. 
It does appear that Cartwright's terminology proposal remained obscure. 
In the book by \cite{guala2005}, we read: ``Genuine \emph{replications} [\ldots] generally involve some (minor or major) modifications of the original design.'' 
And: ``It is necessary to distinguish, however, between the \emph{replication} and mere \emph{repetition} of an experiment. Repetition is the business of doing an experiment again, trying to keep exactly the same design as in the original.'' 
In a footnote: ``The distinction introduced here is well known in the philosophical literature on experiments, although other authors sometimes use another terminology (e.g., `replication' vs. `reproduction').''---citing \cite{cartwright1991}.
Like in political science, the economics literature uses `replication` much more frequently, often as an umbrella term. (A Google Scholar search with ``replication economics'' gives 298,000 results, and with ``reproducibility economics,'' 53,700 results; checked Jan.~19, 2018).
An even older example in the field of economics is \cite{dewaldETal1986}, while a more recent one is \cite{mccullough2008}. They don't distinguish `reproduce` and `replicate.`
The new-section announcement for replication studies in the \emph{Journal of Applied Econometrics} \cite[]{pesaran2003}, meanwhile, explains what they should include: checking consistency and accuracy of data, checking validity of computations either directly or using different software (preferred), and checking ``if the substantive empirical finding of the paper can be replicated using data from other periods, countries, regions, or other entities as appropriate.'' 
Succinctly put by \cite{duvendackETal2017}: 
``we operationalize `replication' as any study whose main purpose is to determine the validity of one or more empirical results from a previously published study.''

The new monograph by \cite{pellizzariETal2017}, derived from a report of a 2015 workshop, aims to illuminate the semantics problems of the field, falling in agreement with the ACM definitions (group $B1$). 
It cites the National Institute of Standards and Technology (NIST) as ``the most authoritative source'' for the adopted lexicon (though NIST has \emph{not} addressed lexicon for reproducible research, specifically).
Citing from page 8: ``Of particular importance is the crucial distinction between reproducibility and replicability. The differences lie mainly in whether every element of a given study must be done in exactly the same way in any effort either to do it again or to confirm original results.  That requirement is basically the ``restrictive'' aspect of replicability. By contrast, reproducibility does not turn so critically on such elements as having identical subjects, methods, and other components; that is, at least one component differs from the original research study.'' 
Table 1 includes the ``NIST definition'' for reproducibility: 
``closeness of the agreement between the results of measurements of the same measurand carried out under changed conditions of measurement.''
This language appears verbatim on page 14 of NIST Technical Note 1297 \cite[]{taylorETal1994}, where it says that it comes directly from the International Vocabulary of Metrology (VIM). 
It is followed by a note explaining that reproducibility can be expressed by dispersion statistics (e.g., standard deviation), because it refers to measuring the \emph{same} physical quantity, with at least one condition changed (e.g., the location, or the operator, or the method of measurement). 
In this sense, I see no contradiction with the usage of `reproducible` by Claerbout and Donoho, where at least operator, location and time are meant to change on reproducing some results. 
The key, though, is that `replication' or `replicability' are \emph{not} defined terms in any NIST document or in the VIM.
\cite{pellizzariETal2017} claim that ``NIST does, however, use the term in specific applications without defining it or distinguishing it from repeatability''---without a citation. 
They do reference the agreement with \cite{casadevall_fang2010}, which in turn cite \cite{drummond2009}, as we saw previously.
The phrase `replicable measurements' appears in both the definitions of repeatability and reproducibility conditions of measurement in the International Vocabulary of Metrology \cite[]{jcgm2008}, as also mentioned.
Finally, we should note that NIST, despite its name, is not a standards developing organization. 
So to speak of ``NIST definitions'' is itself problematic, because they do not, in fact, write standards. They only endorse the standards of other organizations like ANSI, ISO, IEEE, etc.

In group $A$ are the widely cited works of the Open Science Collaboration---two large-scale efforts to replicate published empirical and correlational studies in psychology. They use the terms `reproduce' and `replicate' interchangeably. 
Explicitly, the first work says: ``Some distinguish between `reproducibility' and `replicability' by treating the former as a narrower case of the latter (e.g., computational sciences) or vice versa (e.g., biological sciences). We ignore the distinction.'' \cite[]{openscience2012}. 
In their later work \cite[]{openscience2015}, the authors recurrently use `replication' (208 times), and less frequently apply `reproducible,' as an umbrella term (60 times). 
However, the lead author of these studies, Brian Nosek, adopted the distinction between terms according to Claerbout/Donoho/Peng when speaking in a recent video interview \cite[]{nsf2015vid}. 
He says:
``[by] reproducibility\ldots what we're usually referring to is: can you take the data and findings that I produced in some research, and run the analysis again and get the result back that I got.
[In] replicability\ldots I do a study, get a finding with some data, and you do your own study\ldots''

\cite{stodden2015} also uses the two recurrent terms indistinctly: 
``A published finding may not reproduce in independent replications of the original experimental design (i.e., reimplementing the experiment) for any of several statistical reasons.'' 
This work focuses on \emph{statistical reproducibility}, which ``refers to the failure to replicate an experiment owing to flawed experimental design or statistical analysis.''
Like some others (see below), this work sidesteps the definition tangle using \emph{reproducibility} as an umbrella term, and proposing expanded terminology that distinguishes between \emph{empirical} reproducibility, \emph{computational} reproducibility, and \emph{statistical} reproducibility.

\cite{lejaeghere2016} tackle for the first time the issue of reproducibility in the context of computational chemistry and materials science. 
This is an impressive work involving dozens of researchers, and comparing 15 solid-state codes on 71 elemental crystals, over many thousands of calculations. 
`Reproducibility' here takes the meaning of a new analysis with different methods (i.e., on the `replication' side of Peng's spectrum). 
But their use of the term is generic: ``Scientific results are expected to be reproducible. When the same study is repeated independently, it should reach the same conclusions.''
I have grouped this work with $A$ since the terms `replication/replicate' are absent, and the authors appear unaware of the different terminologies.
A recent editorial in one of the top chemistry journals, on the other hand, plainly aligns with the Claerbout/Donoho sense for `reproducible research` when giving a description of what this means in computational chemistry of materials: open data, using standard machine-readable formats, hosting in archival-quality repositories providing a persistent URL or DOI, fully sharing input and output files, and open-source software \cite[]{coudert2017}.

Hundreds of works could of course be added to this discussion. 
I used no special criteria to choose works to cite here, beyond having come across them in the past, read them, and taken notes about their terminology (plus a random walk through Google Scholar while writing). 

\section*{Expanded terminologies}
\vspace{\up}

\cite{stodden2013} made first a separation between \emph{empirical} reproducibility and \emph{computational} reproducibility, then further unpacked \emph{statistical} reproducibility \cite[]{stodden2015}.
A few others have also invented expanded terminologies. 
In the fields of economics and political science, the umbrella term is \emph{replication}. 
\cite{hamermesh2007} expands this into \emph{pure} replication, \emph{statistical} replication, and \emph{scientific} replication. The first refers to re-analyzing the same data with the same model and estimation parameters; the second may use different but comparable data, but the same model. 
The last one would use ``different sample, different population, and perhaps similar but not identical model [\ldots] and, indeed, comprises most of what economists view as replication.''

In the context of neuroscience, \cite{crookETal2013}---after citing \cite{drummond2009} as their source for terminology---propose to make further distinctions among: 
internal replicability, external replicability, cross replicability, and reproducibility. 
In the first case, the original authors or someone else in the same group can re-create the results, re-executing the same software. In the second case, a reader of published results can re-create them using the data and code supplied by the authors. And cross-replicability would mean running the same model but with different software. But then \cite{crookETal2013} say that ``the boundary line between cross-replicability and reproducibility is not always clear.''

\cite{goodmanETal2016} recognize that ``basic terms---reproducibility, replicability, reliability, robustness, and generalizability---are not standardized.'' 
They propose new lexicon: \emph{methods} reproducibility (the original meaning of reproducibility by Claerbout/Donoho), \emph{results} reproducibility (the meaning of replicability in Peng), and \emph{inferential} reproducibility.

\section*{Beyond the academic literature}

\subsection*{Funding agencies}
\vspace{\up}

The National Science Foundation Subcommittee on Replicability in Science of the Advisory Committee to the NSF Directorate for Social, Behavioral, and Economic Sciences (SBE) published a report titled ``Perspectives on Robust and Reliable Science'' in May 2015 \cite[]{bollenETal2015}.
They define terminology as follows: 
``reproducibility refers to the ability of a researcher to duplicate results of a prior study using the same materials as were used by the original investigator,'' and
``new evidence is provided by new experimentation, defined in the NSF report as `replicability'.''
The report recognizes a lack of consensus on terminology, but adopts clear definitions for the two recurrent terms. 
It also states that ``reproducibility is a minimum necessary condition for a finding to be believable and informative.'' 
These definitions are consistent with the Claerbout/Donoho/Peng convention (group $B1$)---despite the fact that  in economics and political science `replicability` had been widely used as an umbrella term.

The NSF Directorate for Computer \& Information Science \& Engineering (CISE) issued a ``Dear Colleague Letter: Encouraging Reproducibility in Computing and Communications Research'' in October 21, 2016.\footnote{\;\url{https://www.nsf.gov/pubs/2017/nsf17022/nsf17022.jsp}}
It encourages PIs to ``embrace completeness and transparency in developing rigorous protocols as well as in making experimental parameters and collected data available to other researchers','
but it makes no attempt to define `reproducibility.'
It cites the December 2011 special issue of \emph{Science} containing the perspective of \cite{peng2011}, and a May 2016 editorial in \emph{Nature} introducing a reader survey of concern with reproducibility \cite[]{baker20161}; the latter work uses reproduce/reproducibility as umbrella terms.

The National Institutes of Health (NIH) maintains a website titled ``Rigor and Reproducibility,'' but does not provide terminology.\footnote{\;\url{https://grants.nih.gov/reproducibility/index.htm}}
It set up requirements for strengthening reproducibility of biomedical research in January 2016, providing instructions for grant proposals. NIH focuses on \emph{rigor} (concerning methods and materials) and \emph{transparency} (concerning publication). 
Their definition of reproducibility is the straight combination of those two notions, while making no particular distinction with the word `replication' \cite[]{collinsETal2014}.

\subsection*{Professional societies}
\vspace{\up}

The 1999 Ethics and Values document \footnote{\;\url{https://www.aps.org/policy/statements/99_6.cfm}} of the American Physical Society (APS) reads: ``The success and credibility of science are anchored in the willingness of scientists to [\ldots] Expose their ideas and results to independent testing and replication by others. This requires the open exchange of data, procedures and materials.'' 
The use of the word `replication' coincides with the ideas of the Peng spectrum, but it could be a coincidental choice of words.

The American Economic Association (AEA) has a policy\footnote{\;\url{https://www.aeaweb.org/journals/policies/data-availability-policy}} that applies to all its journals: ``to publish papers only if the data used in the analysis are clearly and precisely documented and are readily available to any researcher for purposes of replication.'' 
(This includes all computer codes, configuration files and scripts.) 
They use the term `replication` throughout, but the common sense is the replication of a full study or published paper. 
For example, \cite{coffmanETal2017} propose that journals publish replication reports, which ``summarize novel work replicating an existing high-impact paper, or they could highlight a replication result embedded in a wider-scope published paper.''

The Association for Psychological Science (APS) has been promoting open science and replication efforts for some years. 
In 2012, their journal \emph{Perspectives on Psychological Science} began publishing special sections on replication \cite[]{pashler2012}.
The APS flagship journal, \emph{Psychological Science}, started several initiatives to enhance replicability \cite[]{eich2014}, including: stronger methods reporting (not counted towards word limits), expanded statistical requirements, and new badges for recognizing and promoting open-science practices.
A later editorial \cite[]{lindsay2015} reaffirmed the commitment to replication, while discussing typical practices that weaken it, like low statistical power and $p$-hacking. 
In all these contexts, `replication' is used broadly, but often referring to independent studies aimed and confirming the scientific findings of a previous publication.
The APS is also an original signatory of the Transparency and Openness Promotion (TOP) Guidelines \cite[]{nosekETal2015}. 
These guidelines, signed so far by nearly 3,000 journals and organizations, associate reproducibility to sharing of data, open code, research design disclosure, pre-registration of analysis plans and study details, and replication studies.

The American Psychological Association (APA) just released new journal article reporting standards, with recurrent mention of the phrase `replication studies' to mean complete studies (including data collection) meant to confirm the findings of another  \cite[]{appelbaumETal2018}.

The Federation of American Societies for Experimental Biology published a recommendations document \cite[]{faseb2016} targeted at biomedical and biological research (particularly two key tools: mouse models and antibodies). 
It provides definitions that fall squarely in group $B2$:

\vspace{\up}
\begin{description}
\item[Replicability:] the ability to duplicate (i.e., repeat) a prior result using the same source materials and methodologies. This term should only be used when referring to repeating the results of a specific experiment rather than an entire study;
\item[Reproducibility:] the ability to achieve similar or nearly identical results using comparable materials and methodologies. This term may be used when specific  findings from a study are obtained by an independent group of researchers.
\end{description}

The American Statistical Association (ASA) released in January 2017 a document with recommendations for funding agencies on supporting reproducible research \cite[]{bromanETal2017}.
Its definitions align with group $B1$:

\vspace{\up}
\begin{description}
\item[Reproducibility:] A study is reproducible if you can take the original data and the computer code used to analyze the data and reproduce all of the numerical findings from the study. This may initially sound like a trivial task but experience has shown that it?s not always easy to achieve this seemingly minimal standard.
\item[Replicability:] This is the act of repeating an entire study, independently of the original investigator without the use of original data (but generally using the same methods).
\end{description}

\subsection*{Other journal policies and editorials}
\vspace{\up}

The \emph{Journal of Applied Econometrics} (JAE) has maintained a data archive\footnote{ \url{http://qed.econ.queensu.ca/jae/}} since the late 1980s, making it mandatory for papers published after January 1994 to deposit their (non-confidential) data. 
Depositing software, however, is voluntary.
JAE announced in \cite{pesaran2003} a new section dedicated to replications of previously published empirical results. 
This journal is the leader in publishing such works among journals in the field, followed by the \emph{American Economic Review}, according to the survey of \cite{duvendackETal2017}.

The \emph{American Journal of Political Science} requires authors\footnote{ \url{https://ajps.org/ajps-replication-policy/}} to ``provide replication materials that are sufficient to enable interested researchers to reproduce all of the analytic results that are reported in the text and supporting materials.'' 
Their terminology does not distinguish between the two recurrent terms (group $A$), 
but their use of `replication files'---meaning, all data and code that produce the results presented in an accepted paper---corresponds to the yardstick of Claerbout and Donoho's `reproducible research.' 
What is remarkable about this journal's process is that they contract a third party\footnote{~The Odum Institute for Research in Social Science, in the case of quantitative analyses, and the Qualitative Data Repository at Syracuse University, in the case of qualitative analyses.} to complete a full verification that the author-provided data and analysis code do produce the reported results.
They make this a condition for publication. 
Author-provided files are deposited in the journal's section on the Dataverse repository,\footnote{\,\url{https://dataverse.harvard.edu/dataverse/ajps}} where they are labeled `Replication Data for <paper title>.'

The statement of the \emph{Annals of Internal Medicine} \cite[]{laineETal2007} talks about `replication by independent scientists in independent settings' and `reproducible research,' when published results can be re-created using the procedures and data used by the original authors. 
This language is consistent with that of group $B1$.

The journal \emph{Biostatistics} announced its reproducible-research initiative in \cite{peng2009}, with a clear distinction between replication of scientific findings and the minimum standards of reproducible research. It designates articles as reproducible when three conditions are met: open data, open-source code, and the associate editor for reproducibility was able to re-run the analysis to recreate the reported results.

Adding to an existing firm open-data policy, the journal \emph{Genome Biology} recently announced an open-source code requirement.
The editorial reads: ``We consider source code availability and accessibility to be important to ensure that computational analyses can be easily repeated and research shown to be reproducible.'' \cite[]{marszalek2016}. 
A more recent piece advocating for structured supplemental materials submits that an entire workflow could be   included in supplements: ``Workflows are especially relevant for in silico analyses, as reproducibility can turn on the ability to recreate the exact parameters employed.'' \cite[]{greenbaumETal2017}. 
This sense is similar to Claerbout and Donoho's idea of reproducible research, while the term `replication` is hardly used in this field, where the word is more likely to  appear in phrases like `DNA replication' and `genome replication.'

I also want to mention the new \emph{ReScience} journal:\footnote{\,\url{https://rescience.github.io}} ``a peer-reviewed journal that targets computational research and encourages the explicit replication of already published research.'' 
As described in their FAQs, a publication in \emph{ReScience} achieves two goals: 
``by replicating the original work, it provides an independent implementation of the original computational protocol, and by making this new implementation public, it is reproducible and thus a safer basis for future research to build on.'' 
A recent paper reporting on the initiative's progress  gives the adopted terminology \cite[]{rougierETal2017}:
\vspace{\up}
\begin{description}
\item[Reproducing]the result of a computation means running the same software on the same input data and obtaining the same results. The goal of a reproduction attempt is to verify that the computational protocol leading to the results has been recorded correctly.
\item[Replicating]a published result means writing and then running new software based on the description of a computational model or method provided in the original publication, and obtaining results that are similar enough to be considered equivalent.
\end{description}

\section*{Conclusions}
\vspace{\up}

The Claerbout/Donoho/Peng terminology is broadly disseminated across disciplines (see Table 2). 
But the recent adoption of an opposing terminology by two large professional groups---ACM and FASEB---make standardization awkward. 
The ACM publicizes its rationale for adoption as based on the International Vocabulary of Metrology, but a close reading of the sources makes this justification tenuous. 
The source of the FASEB adoption is unclear, but there's a chance that \cite{casadevall_fang2010} had an influence there. They, in turn, based their definitions on the emphatic but essentially flawed work of \cite{drummond2009}.

\begin{table}[h]
\caption{Grouping of terminologies, as in Table 1, but by discipline.}
\begin{footnotesize}
\begin{center}
\begin{tabular}{c c c}
$A$				&	$B1$				&	$B2$ \\ \hline
political science		& signal processing		& microbiology, immunology (FASEB) \\
economics			& scientific computing	& computer science (ACM) \\
				& econometry & \\
				& epidemiology & \\
				& clinical studies & \\
				& internal medicine & \\
				& physiology (neuro) & \\
				& computational biology & \\
				& biomedical research & \\
				& statistics & \\

\end{tabular}
\end{center}
\end{footnotesize}
\label{repro-table}
\end{table}%

It seems that any attempt to unify terminologies can only start by conversations with ACM and FASEB to explore how slim the chance might be that they backtrack their adoption to align with the predominant usage.
Conversations with leaders in the fields of political science and economics, where `replication` is  predominant, should follow. 
Use of the term here often applies to a new study that aims to confirm findings reported in a publication (and thus aligns with Peng's), but it also spills over to calling the research compendia `replication files.' 
This custom in the \emph{American Journal of Political Science} is unlikely to be reversed, but it can be absorbed into the Claerbout/Donoho/Peng convention, as reproducible research is also necessary for successful replication. 
Whereas economics and political science are predominantly empirical disciplines, using software for statistical analysis is ubiquitous. 
A feasible way out would be if these fields accept calling `reproducible research' that which openly archives full sets of `replication files.'


{\small
\bibliographystyle{jponew}

}

\end{document}